\documentclass[lettersize,journal]{IEEEtran}
\usepackage{bm}
\usepackage{amsmath,amsfonts}
\usepackage{algorithmic}
\usepackage{algorithm}
\usepackage{array}
\usepackage[caption=false,font=normalsize,labelfont=sf,textfont=sf]{subfig}
\usepackage{textcomp}
\usepackage{stfloats}
\usepackage{url}
\usepackage{verbatim}
\usepackage{graphicx}
\usepackage{cite}
\usepackage{gensymb}
\usepackage{xcolor}
\usepackage{colortbl}
\usepackage{adjustbox}
\usepackage{multirow}
\usepackage{array} 
\usepackage{tabularray}
\usepackage{balance}
\begin{document}

\title{Theoretical and Empirical Study of Spatial Power Focusing Effect for Sparse Arrays at Terahertz Band}
%{Channel Measurements and Modeling in Meeting room for virtual MIMO from 260 to 400 GHz}
\author{Yongchao He, Taihao Zhang, Cunhua Pan,~\IEEEmembership{Senior Member,~IEEE}, Hong Ren,~\IEEEmembership{Member,~IEEE}, Xianzhe Chen,\\ Tian Qiu, Bingchang Hua, Jiangzhou Wang,~\IEEEmembership{Fellow,~IEEE}
        % <-this % stops a space
\thanks{This work was supported by the Key Research and Development Projects under Grant 2023YFB2905100. Y. He, T. Zhang, C. Pan, H. Ren, X. Chen, T. Qiu and J. Wang are with the National Mobile Communications Research Laboratory, School of Information Science and Engineering, Southeast University, Nanjing 211189, China (e-mail: {heyongchao, taihao, cpan, hren, chen.xianzhe, tianqiu, j.z.wang}@seu.edu.cn). B. Hua is with Institution: Purple Mountain Laboratories, Nanjing, Jiangsu 211111, China (e-mail: huabingchang@pmlabs.com.cn).

\textit{Corresponding author: Cunhua Pan.}
 }% <-this % stops a space
%\thanks{Manuscript received April 19, 2021; revised August 16, 2021.}
}



\maketitle

\begin{abstract}
 This work investigates the spatial power focusing effect for large-scale sparse arrays at terahertz (THz) band, combining theoretical analysis with experimental validation. Specifically, based on a Green's function channel model, we analyze the power distribution along the $z$-axis, deriving a closed-form expression to characterize the focusing effect. Furthermore, the factors influencing the focusing effect, including phase noise and positional deviations, are theoretically analyzed and numerically simulated. Finally, a 300 GHz measurement platform based on a vector network analyzer (VNA) is constructed for experimental validation. The measurement results demonstrate close consistence with theoretical simulation results, confirming the spatial power focusing effect for sparse arrays.
\end{abstract}


\section{Introduction}
\IEEEPARstart{T}{erahertz} (THz) communication, regarded as a key technology for 6G systems, offers ultra-wide bandwidth for extremely high data rates \cite{THz,THz2}.
To overcome the severe propagation loss at such frequencies, base stations (BS) are required to equip extremely large-scale arrays (XL-arrays), which significantly extend the near-filed (Fresnel) region to hundreds of meters \cite{XLMIMO}. 
Consequently, high-frequency communications are expected to operate in the near field, where spherical wavefronts
enable superior spatial resolution and precise beam focusing capabilities, surpassing what is achievable in the conventional far field \cite{near}.
However, densely spaced XL-arrays with half-wavelength antenna spacing require a massive number of elements,  leading to prohibitive hardware costs and power consumption \cite{duo}. Sparse arrays emerge as a promising alternative to overcome these limitations \cite{sparse}. It is thus imperative to explore sparse arrays in the near field at THz frequencies.

Considerable attention has been devoted to large-scale sparse arrays in prior research, particularly their unique spatial and energy distribution characteristics. 
For instance, the work \cite{another} integrated massive multiple-input multiple-output (MIMO) architectures with sparse array systems, demonstrating the potential to mitigate mutual coupling, increasing degrees of freedom (DoF), and improving spatial resolution. Recently, \cite{nanfang} investigated the near-field beamspace patterns of two sparse array configurations, namely the linear sparse array (LSA) and extended coprime array (ECA), and developed a hybrid beamforming design scheme.
Furthermore, \cite{chen} provided closed-form analysis of effective degrees of freedom (EDoF) in the near-field XL-MIMO system with sparse arrays, and revealed that larger antenna spacing can reduce interference caused by the main lobe and side lobes. It identified a power focusing effect, referring to  the ability of sparse arrays to concentrate electromagnetic energy at a specific spatial location, thereby significantly enhancing the signal strength on the focal point.

In this paper, we present comprehensive analysis of the key factors influencing the spatial power focusing effect, including phase noise as well as positional deviations. To experimentally validate the theoretical analysis, we further establish a vector network analyzer (VNA)-based measurement platform to conduct experimental measurements at 300 GHz. 
The experimental results corroborate the theoretical analysis and demonstrate that the focusing effect exhibits remarkable sensitivity to positional deviations while maintaining strong robustness against phase noise. These characteristics underscore its promising potential for high-precision sensing and alignment in practical applications.

\section{System Model}\label{model}

\begin{figure}
	\centering
	\includegraphics[width=7.5cm]{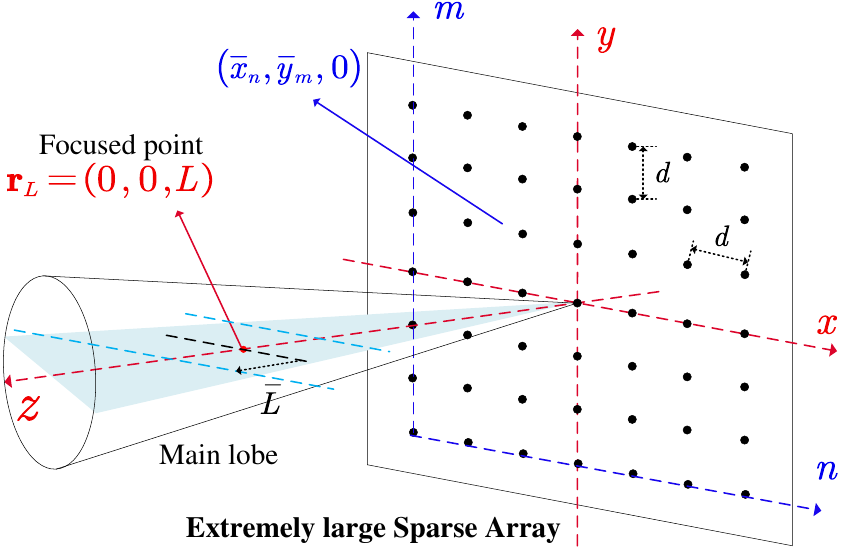}
	\caption{Schematic illustration of the sparse UPA system.}
	\label{figsys} 
\end{figure}

Consider a downlink near-field communication system comprising a sparse $\sqrt{N}$ $\times$ $\sqrt{N}$ uniform planar array (UPA) deployed at the BS for signal transmission to users, as illustrated in Fig. \ref{figsys}. For analytical tractability, only the line-of-sight (LoS) path is considered, with any obstacles or reflections neglected. The UPA is positioned on the $XY$ plane, with its edges aligned with the coordinate axes. The antenna elements are uniformly spaced in both the horizontal and vertical directions with the identical spacing of $d$. $L$ represents the distance between the point $\mathbf{r}_L$ and the center of the UPA. $\bar{L}$ denotes the displacement along the $z$-axis from point $\mathbf{r}_L$. 

The received signal $g_L$ at the location $\mathbf{r}_L=\left( 0,0,L\right)$ can be expressed as
\begin{align}
g_L = \sum_{n = 1}^{\sqrt{N}}\sum_{m = 1}^{\sqrt{N}}h_{n,m}^{L}e^{i\varphi_{n,m}^{L}},
\end{align}
where $n$ and $m$ denote the antenna indices along the horizontal and vertical 
directions of the UPA, respectively, as depicted in Fig. \ref{figsys}. Here, the $h_{n,m}^{L}$ denotes the channel coefficient between the $\left( n,m \right)$-antenna and the point $\mathbf{r}_L$, generated from the scalar Green’s function. As derived in \cite{2019Waves}, the scalar Green’s function between an arbitrary transmitting point $\mathbf{s}$ and an arbitrary receiving point $\mathbf{r}$ can be given by
\begin{align}
\mathbf{G}(\mathbf{r},\mathbf{s})=-\frac{e^{(ik\left\| \mathbf{r}-\mathbf{s} \right\| )}}{4\pi \left\| \mathbf{r}-\mathbf{s} \right\|},
\end{align}
where $k=2\pi/\lambda$ is the wavenumber, and $\lambda$ is the wavelength. 
Due to the fact that the power variations across the array are negligible in the radiative near field compared to the phase variations \cite{apro}, the approximation can be justified. Thus, the channel coefficient $h_{n,m}^{L}$ is calculated as 
\begin{align}
h_{n,m}^{L}&=-\frac{e^{ik\left\| (x_n,y_m,0)-(0,0,L) \right\|}\lambda}{4\pi \left\| (x_n,y_m,0)-(0,0,L) \right\|}\nonumber
\\
&\approx -\frac{e^{i\frac{2\pi}{\lambda}\sqrt{x_{n}^{2}+y_{m}^{2}+L^2}}\lambda}{4\pi  L },
\end{align}
where $x_n$ and $y_m$ denote the horizontal and vertical coordinates given by
\begin{align}
x_n=\left( n-\frac{\sqrt{N}+1}{2} \right) d,\,y_m=\left( m-\frac{\sqrt{N}+1}{2} \right) d.
\end{align}
The term $\varphi _{n,m}^{L}$ denotes the phase shift applied to antenna element $\left( n,m \right)$, which is configured via the precoding vector to steer the beam toward the focal point $\mathbf{r}_L$, with the factor $\frac{1}{\sqrt{N}}$ applied for normalization. Assuming that the total transmit power of the BS antenna array is $P$, the received power of $\mathbf{r}_l=\left( 0,0,l\right)$ can be written as
\begin{align}\label{P0}
P_l&=P\left| \sum_{n=1}^{\sqrt{N}}{\sum_{m=1}^{\sqrt{N}}{h_{n,m}^{l}\frac{1}{\sqrt{N}}e^{i\varphi _{n,m}^{L}}}} \right|^2 \nonumber
\\
&=\beta _l\left| \sum_{n=1}^{\sqrt{N}}{\sum_{m=1}^{\sqrt{N}}{e^{i\frac{2\pi}{\lambda}D^{l}_{n,m}} e^{i\varphi _{n,m}^{L}}}} \right|^2, 
\end{align}
where $l=L+\bar{L}$, $\beta _l=\frac{P}{N\left( 4\pi l \right) ^2}$, and $D_{n,m}^l=\sqrt{x_{n}^{2}+y_{m}^{2}+l^2}$ represents the distance between the $\left( n,m \right)$-th antenna element and $\mathbf{r}_l$.
To achieve beam focusing at $\mathbf{r}_L$, the precoding vector is designed such that the received power $P_L$ at this point is maximized. Therefore, the phase shift $\varphi _{n,m}^{L}$ satisfies
\begin{align}
\varphi _{n,m}^{L}=-\frac{2\pi}{\lambda}\sqrt{x_{n}^{2}+y_{m}^{2}+L^2}.
\end{align}

\section{Spatial Power Focusing Effect}
\begin{figure}[h]
	\centering
	\includegraphics[width=8.5cm]{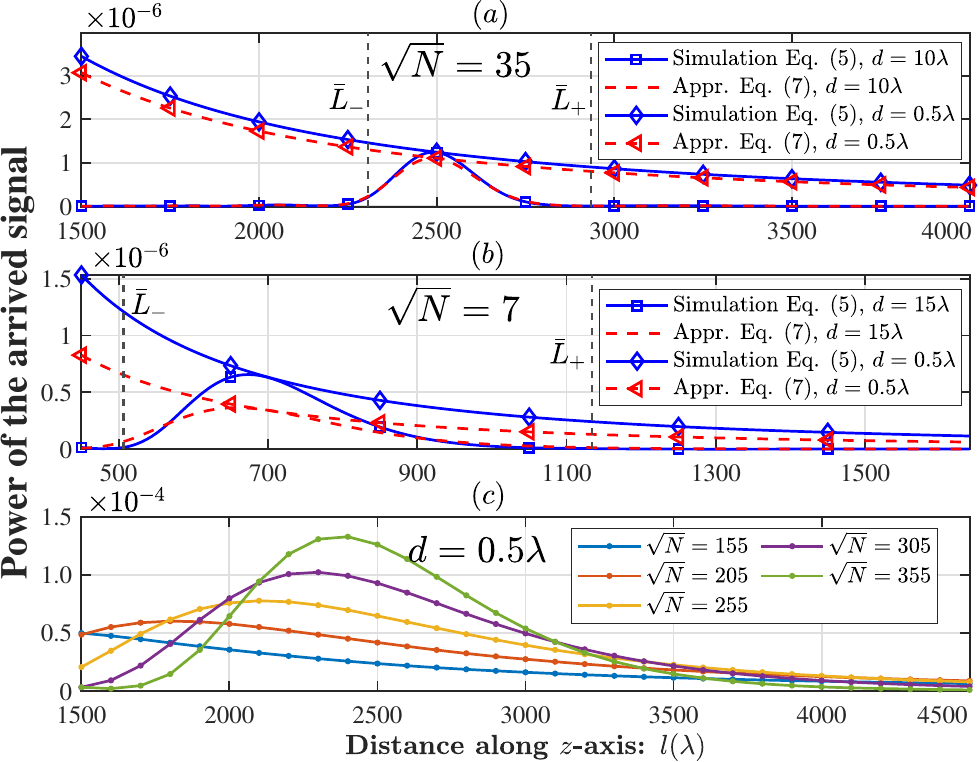}
	\caption{Signal power along $z$-axis with different $\sqrt{N}$ and $d$.}
	\label{figP0} 
\end{figure}

When a sparse array system equipped with UPAs transmits a signal toward the focal point $\mathbf{r}_L$ in the near field, the distribution of received power along the $z$-axis differs significantly from that of dense arrays with half-wavelength spacing. As shown in Fig. \ref{figP0}, a preliminary simulation of $P_L$ was conducted based on Eq. \eqref{P0}. The simulation was carried out at 300 GHz with a transmit power of $P=1$ W, using the following configurations: (a) $\sqrt{N}=35$, $L=2500\lambda$, $d=\left\{ 10\lambda ,0.5\lambda \right\}$; (b) $\sqrt{N}=7$, $L=700\lambda$, $d=\left\{ 15\lambda ,0.5\lambda \right\}$; (c) different $\sqrt{N}$, $L=2500\lambda$, $d=0.5\lambda$. 

As indicated by the blue curve of Fig. \ref{figP0}-(a), (b), for the case of $d=0.5\lambda$, the received signal power decreases monotonically as $l$ increases. In contrast, the sparse array with larger antenna spacing, operating in the near field, concentrates the received power at the focal point. It is characterized by a prominent peak in the received power along the $z$-axis after beamforming.
However, for dense arrays with half-wavelength spacing, as in Fig. \ref{figP0}-(c), achieving the focusing effect requires a substantially large number of antennas $N$ \footnote{In the near field, the array gain of an XL-UPA saturates with roughly $10^{6}$ antenna elements \cite{zhi}, and our setup in Fig. \ref{figP0}-(c) operates within this regime. Thus other constraints need not be considered.}, which leads to significant cost and implementation challenges.
%It is observed that increasing the number of antennas leads to a more pronounced spatial power focusing effect, which is characterized by a prominent peak in the received power along the $z$-axis after beamforming. However, both simulation and analysis indicate that achieving a distinct spatial power focusing effect with a conventional half-wavelength UPA demands a number of antennas far exceeding practical configurations.
%Therefore, we investigate a sparse array architecture that featurs uniform but substantially increased antenna spacing to achieve comparable focusing performance with fewer elements.
Hence, we consider sparse arrays to achieve beam focusing effect. 
%Moreover, it can be observed from the Fig. \ref{figP0}-(c) that the received power along the $z$-axis decreases as the number of antennas increases. This occurs because the spatial power focusing effect introduces unexpected side lobes \cite{chen}, which disperse part of the energy away from the main lobe under a fixed total transmit power.

In previous work \cite{chen}, it conducted an initial investigation of this phenomenon, presenting a derived expression for the received signal.
%Regarding the focusing effect of sparse arrays, the detailed theoretical analysis was conducted in our previous work \cite{chen}, where a closed-form expression for the received signal was derived. 
Specifically, when a sparse array focuses on the point $\mathbf{r}_l$, the power $P_{l}$ at the location $\mathbf{r}_{l}=\left( 0,0,l \right)$ can be approximately expressed as \cite{chen}
\begin{align}\label{power}
P_l\approx \frac{P\lambda^2}{\left( 4\pi  l \right) ^2}\cdot \rho _{\bar{L}},
\end{align}
where $\rho _{\bar{L}}$ is given by
\begin{align}
\rho _{\bar{L}}\approx \begin{cases}
	\frac{\left( \sqrt{N}-1 \right) ^4}{Nb^4}\left( C^2\left( b \right) +S^2\left( b \right) \right) ^2,\eta \ne 0\\
	\frac{\left( \sqrt{N}-1 \right) ^4}{N},\eta =0\\
\end{cases},
\end{align}
\begin{align}\label{yita}
b=\sqrt{\left| \frac{\pi d^2}{\lambda L} \eta\right|} \cdot \frac{\sqrt{N}-1}{2}, \,\,\eta =\frac{\bar{L}}{L+\bar{L}}.
\end{align}
Functions $C(\cdot)$ and $S(\cdot)$ are the Fresnel integrals, given respectively by
$C\left( x \right) =\int_0^x{\cos \!\left( \frac{\pi}{2}t^2 \right)}dt$ and $S\left( x \right) =\int_0^x{\sin \!\left( \frac{\pi}{2}t^2 \right)}dt$.

In Eq. \eqref{power}, the first factor reflects the power variation due to path loss, and the second factor $\rho _{\bar{L}}$ accounts for the variation induced by the phase variation across the UPA. From \cite{chen}, the length of the main lobe of the arrived signal along the $z$-axis centered at $\mathbf{r}_L=\left( 0,0,L \right)$ can be written as
\begin{align}
\left| \eta \right|=\frac{4b_{min}^{2}\lambda L}{\left( \sqrt{N}-1 \right) ^2\pi d^2},
\end{align}
where $b_{min}=1.9111$ denotes the local minimum point closest to $b = 0 $. Based on Eq. \eqref{yita}, the length of the main lobe in $z_+$-axis and $z_-$-axis can be obtained as
\begin{align}
\bar{L}_+=\frac{L}{\frac{\pi d^2\left( \sqrt{N}-1 \right) ^2}{4b_{min}^{2}\lambda L}-1},\,\bar{L}_-=\frac{-L}{\frac{\pi d^2\left( \sqrt{N}-1 \right) ^2}{4b_{min}^{2}\lambda L}+1}.
\end{align}
Based on the above theoretical derivations, we conduct simulations under identical configurations, as indicated by the red dashed curves in Fig. \ref{figP0}. The red dashed lines align closely with the blue solid lines. This result theoretically reaffirms the focusing effect of sparse arrays, thereby laying the groundwork for subsequent experimental validation.

\section{ Experimental Verification at Terahertz Frequencies}
To validate the spatial power focusing effect, experiments are conducted using a VNA-based measurement platform. Then, we examine the factors influencing this effect, including phase noise and positional deviations.

\subsection{Experiment setup}
\begin{figure}
	\centering
	\includegraphics[width=7.5cm]{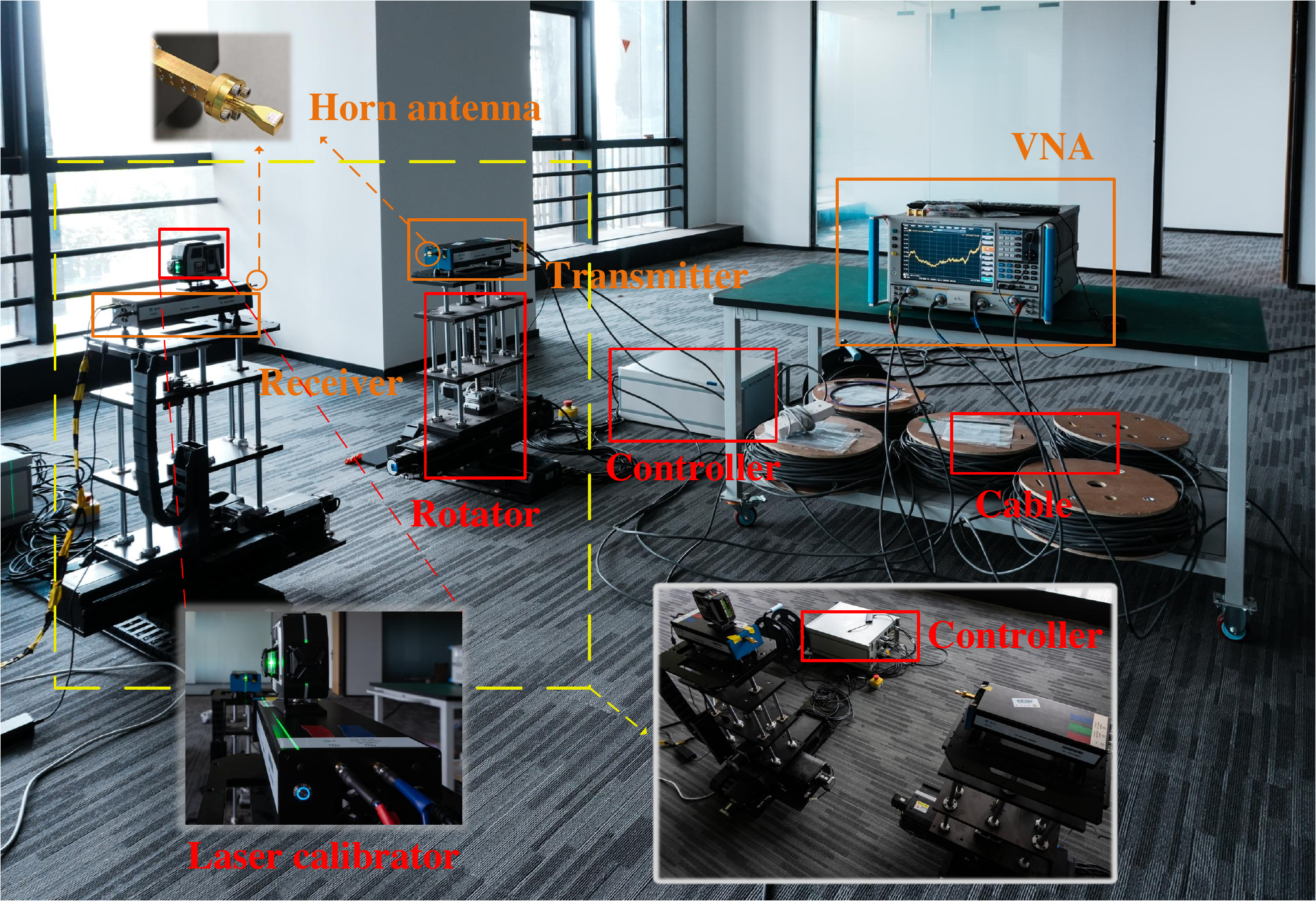}
	\caption{ Experimental setup of the 300 GHz sparse array based on a VNA.}
	\label{figreal} 
\end{figure}

As shown in Fig. \ref{figreal}, the experiment setup comprises the following key components.
\begin{enumerate}
    \item A transmitter (Tx) and receiver (Rx) module, each equipped with a horn antenna featuring a half-power beamwidth of 10$\degree$ and supporting a wide frequency band from 260 to 400 GHz.
    \item A VNA covering the range of 9.62$-$14.81 GHz.
    \item Two rotators, each individually driven by a dedicated controllers.
    \item Five low-loss cables for interfacing the units.
    \item A laser calibrator used for the alignment of the transmit and receive antenna arrays.
\end{enumerate}
We ingeniously construct a virtual UPA using rotator-based automated control. In the setup, the transmitter and receiver modules are mounted on individual rotators, functioning as the Tx and Rx, respectively, with their aperture centers precisely aligned via a laser calibrator. Both the Tx and Rx operate at 300 GHz. A UPA is deployed at the Tx side in the THz sparse arrays system, equipped with $N = 7 \times 7$ antenna elements. The antenna spacing is set to 15 $\lambda$, with the array center located at the origin $(0,0,0)$, and the focal point is fixed at $(0,0,700\lambda)$ to ensure it lies within the near field.

\subsection{Experiment Results}
During the measurements, channel coefficients are obtained using the close-in (CI) path loss model, 
\begin{align}\label{CI}
\mathrm{PL}\left[ \mathrm{dB} \right] =10\cdot \mathrm{PLE}\cdot \lg\left( \frac{l}{d_0} \right) +\mathrm{FSPL}\left( d_0 \right) +X_{\sigma},
\end{align}
where the path loss exponent (PLE) is 1.91 \cite{AWPL}, the reference distance $d_0$ is 1 m for the indoor scenario, $\mathrm{FSPL}$ denotes the free space path loss, and $X_{\sigma}$ represents log-normal shadow fading. To account for the discrepancy\footnote{The channel coefficients from the Green’s function-based model exhibit a scaling discrepancy compared to those from the CI model, primarily due to differences in normalization and modeling assumptions.} with the Green’s function-based model, the channel coefficients are normalized in the subsequent analysis. 
\begin{figure}[h]
	\centering
	\includegraphics[width=7.5cm]{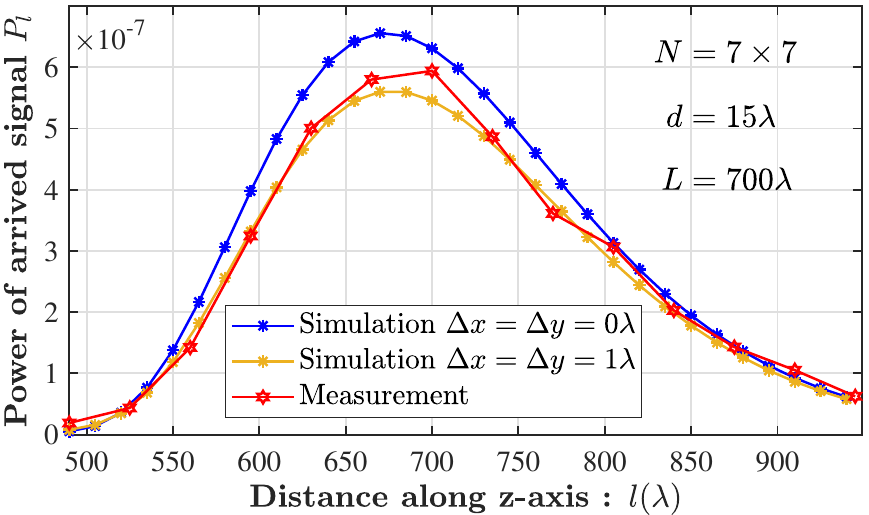}
	\caption{The measurement and simulation of $P_l$ along $z$-axis.}
	\label{figresult} 
\end{figure}

Fig. \ref{figresult} illustrates the power distribution of the arrived signal
along the $z$-axis with $\Delta x = \Delta y = 0$ and $\Delta x = \Delta y = \lambda$, together with the experimental measurements for validation. Here, $\Delta x$ and $\Delta y$ denote the positional deviations of the center antenna element at the Tx side in the horizontal and vertical directions, respectively.
As observed, the measured results exhibit a spatial power focusing effect that fits well with theoretical predictions, thereby validating the proposed sparse arrays model in the near field. Slight differences between measurements and simulations can be primarily attributed to system noise and alignment inaccuracies, which will be analyzed in the following subsection.
%These practical imperfections, together with their theoretical implications, will be analyzed in detail in the following section.
%As observed, the measured results exhibit a focusing effect and fit well with the theoretical simulations even when alignment errors are accounted for. This provides strong validation for the focusing effect of sparse arrays in the near field. 

%Furthermore, during practical measurements, the noise is inevitable, and laser calibrator is employed to mitigate beam misalignment. These factors are likely to influence the focusing performance. 
%Therefore, a comprehensive analysis of the factors is critical for the future application of this effect.

%We can observe that the experimental measurements results exhibit a distinct power focusing phenomenon along the $z$-axis, which indicates the correctness of the theoretical framework. 

%Meanwhile, considering the inevitable minor positional deviations in practical experiments, the simulation curve incorporating such deviations fits well with the measurement results, thereby confirming the focusing effect.

\subsection{Theoretical Analysis of Influencing Factors}
%A comprehensive analysis of the factors influencing the focusing effect for sparse arrays is crucial for guiding experimental validation and future applications of this effect.
\subsubsection{Phase Noise Factor}
\begin{figure}
	\centering
	\includegraphics[width=7.5cm]{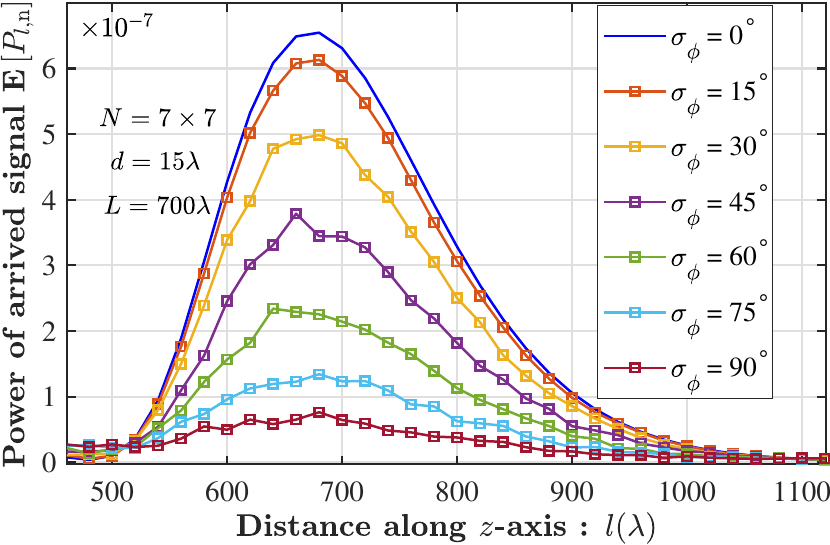}
	\caption{$\mathbb{E} \left[ P_{l,\mathrm{n}}\right]$ along $z$-axis under different $\sigma_\phi^2$.}
	\label{fign} 
\end{figure}

Considering the inevitable presence of noise in practical communication systems, which is clearly reflected in experiment measurement, we reformulate the received power expression Eq. \eqref{P0} as,
\begin{align}\label{17}
P_{l,\mathrm{n}}=\beta _l\left| \sum_{n=1}^{\sqrt{N}}{\sum_{m=1}^{\sqrt{N}}{e^{i\left( \frac{2\pi}{\lambda}D^{l}_{n,m}+\phi _{n,m} \right)}e^{i\varphi _{n,m}^{L}}}} \right|^2. 
\end{align}
Here, the term $\phi_{n,m}$ represents the phase noise following the Gaussian distribution, with variances $\sigma_\phi^2$, applied per antenna element. Thus, the expectation of received power can be expressed as (see the Appendix for the derivation)
\begin{align}\label{18}
\mathbb{E} \left[ P_{l,\mathrm{n}} \right] &=\beta _l\mathbb{E} \left[ \left| \sum_{n=1}^{\sqrt{N}}{\sum_{m=1}^{\sqrt{N}}{e^{i(\frac{2\pi}{\lambda}D_{n,m}^{l}+\phi _{n,m}+\varphi _{n,m}^{L})}}} \right|^2 \right] \nonumber
\\
&=\mu ^2P_l+\beta _lN\left( 1-\mu ^2 \right),
\end{align}
where $\mu$, referred to as the coherence factor, quantifies the power attenuation due to phase noise, given by
\begin{align}\label{19}
 \mu =  \mathbb{E}\left [e^{i\phi_{n,m}}\right]=e^{-\frac{\sigma_\phi^2}{2}}.
\end{align}
We further examine the influence of varying phase noise standard deviations on the received power along the $z$-axis. Simulation results, provided in Fig. \ref{fign}, demonstrate that phase noise degrades the focusing effect, manifesting as a reduction in the average received power, whereas the spatial profile of power focusing remains consistent. This behavior occurs because an increase in $\sigma_\phi^2$ primarily attenuates the coherent component of the signal without altering the spatial distribution dictated by $P_{l,\mathrm{n}}$. Thus, the focusing effect of the sparse array is robust against noise.
\begin{figure}
	\centering
	\includegraphics[width=7.5cm]{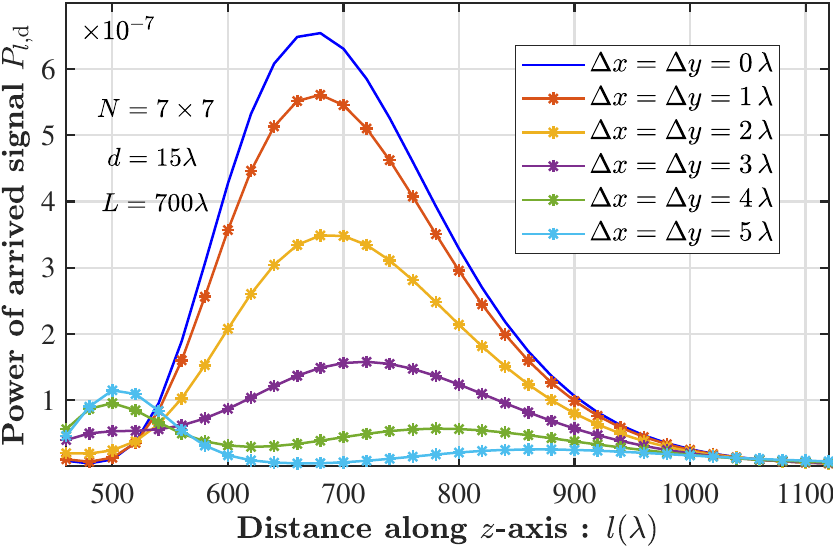}
	\caption{$ P_{l,\mathrm{d}} $ along $z$-axis under positional deviations.}
	\label{figweizhi} 
\end{figure}
\subsubsection{Positional Deviation Factor}
Due to the short wavelengths at high frequencies, positional deviation also becomes a critical factor influencing the focusing effect in practical systems.
The position deviation can be equivalently modeled as a phase error. According to Eq. \eqref{P0}, the received power accounting for the positional deviation at the center antenna element of the UPA is expressed as
\begin{align}\label{Pd}
P_{l,\mathrm{d}}=\beta _l\left| \sum_{n=1}^{\sqrt{N}}{\sum_{m=1}^{\sqrt{N}}{e^{i\frac{2\pi}{\lambda}\bar{D}^{l}_{n,m}}e^{i\varphi _{n,m}^{L}}}} \right|^2,
\end{align}
with the perturbed distance defined by
\begin{align}\label{D}
\bar{D}^{l}_{n,m}=\sqrt{\left( x_n+\Delta x \right) ^2+\left( y_m+\Delta y \right) ^2+l^2},
\end{align}
To evaluate the impact of these deviations, a first-order Taylor series expansion around $\left( \Delta x,\Delta y \right) =\left( 0,0 \right)$ is applied, yielding
\begin{align}\label{22}
\bar{D}^{l}_{n,m}\approx D^{l}_{n,m}+\frac{x_n}{D^{l}_{n,m}}\Delta x+\frac{y_m}{D^{l}_{n,m}}\Delta y,
\end{align}
Higher-order terms $\Delta x^2,\Delta y^2$ are negligible, justified by $ \left| \Delta x \right | , \left| \Delta y \right |\ll D_{n,m}^l$. The resulting phase shift due to positional deviation is therefore approximated as
\begin{align}\label{23}
\Delta \varphi _{n,m}\approx \frac{2\pi}{\lambda}\left( \frac{x_n}{D^{l}_{n,m}}\Delta x+\frac{y_m}{D^{l}_{n,m}}\Delta y \right),
\end{align}
Thus, substituting Eq. \eqref{23} and Eq. \eqref{22} into Eq. \eqref{Pd}, we have
\begin{align}\label{add}
P_{l,\mathrm{d}}\approx\beta _l\left| \sum_{n=1}^{\sqrt{N}}{\sum_{m=1}^{\sqrt{N}}{e^{i\frac{2\pi}{\lambda}D^{l}_{n,m}}e^{i\varphi _{n,m}^{L}}e^{i\Delta \varphi _{n,m}}}} \right|^2.
\end{align}

Fig. \ref{figweizhi} depicts the received power distribution along the $z$-axis under varying positional deviations $\Delta x$ and $\Delta y$. It can be observed that even millimeter-level deviations severely degrade the focusing effect, resulting in a pronounced reduction in received power of sparse terahertz near-field arrays. This degradation is primarily attributable to phase mismatch, which substantially deteriorates the focusing gain. 

Considering only the received power at the focal point $\mathbf{r}_L=\left(0,0,L\right)$, Eq. \eqref{add} can be simplified to
\begin{align}\label{24}
P_{L,\mathrm{d}}\approx\frac{P_L}{N^2}\left| \sum_{n=1}^{\sqrt{N}}{\sum_{m=1}^{\sqrt{N}}{e^{i\Delta \varphi _{n,m}}}} \right|^2. 
\end{align}

\begin{figure}[h]
	\centering
	\includegraphics[width=8cm]{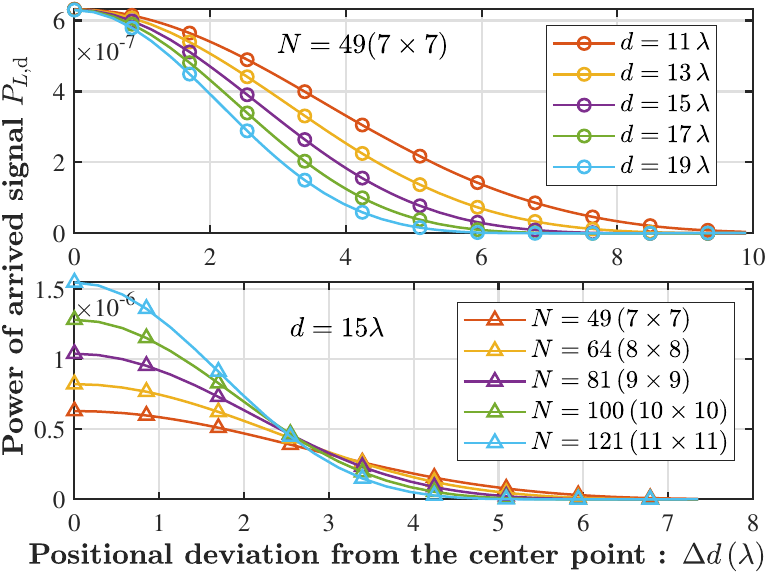}
	\caption{$P_{L,\mathrm{d}}$ vs $\Delta d$ different $\left\{ d,N \right\}$ under positional deviations.}
	\label{figND} 
\end{figure}

The trends in received power as a function of positional deviation for different antenna numbers and spacings are illustrated in Fig. \ref{figND}. It indicates that the attenuation due to positional deviation becomes more pronounced as the number of antennas increases. Similarly, the larger antenna spacings lead to a more rapid decline in received power. This trend can be explained by Eq. \eqref{23}, which shows that the phase sensitivity to positional deviation is proportional to the position of the antenna elements $x_n$ and $y_m$. Therefore, the focusing effect of the sparse array is highly sensitive to positional deviations.

\section{Conclusion}
In this paper, we presented a comprehensive investigation of a discovered spatial power focusing effect for sparse arrays in THz near-field region. The study encompassed both theoretical derivation and experimental validation. We theoretically analyzed factors influencing this effect, including noise and positional deviations. Furthermore, a VNA-based measurement was conducted to experimentally confirm the focusing effect. 

While the full practical potential of this discovered effect remains to be fully explored, our findings suggest it has significant implications. Due to the combination of robustness against noise and exceptional sensitivity to positional deviation, it is particularly suitable for applications demanding ultra-high precision within limited ranges.
Promising examples include its use as a virtual electromagnetic probe for non-destructive testing at microscopic scales, or as a nanoscale alignment and sensing tool in semiconductor manufacturing and micro-system assembly, etc.

\appendix
Considering the noise in practical communication systems, Eq. \eqref{P0} can be rewritten as
\begin{align}
P_{l,\mathrm{n}}=\beta _l\left| S \right|^2,
\end{align}
with $S$ defined as
\begin{align}
&S=\sum_{n=1}^{\sqrt{N}}{\sum_{m=1}^{\sqrt{N}}{e^{i\left( \alpha _{n,m}+\phi _{n,m} \right)}}},\\
&\alpha _{n,m}=\frac{2\pi}{\lambda}D^{l}_{n,m}+\varphi _{n,m}^{L}. 
\end{align}
Then, we further calculate $S$ as follows
\begin{align}\label{24}
\mathbb{E} [ \left| S \right|^2 ] =\sum_{n,m}{\sum_{n\prime ,m\prime}{e^{i\left( \alpha _{n,m}-\alpha _{n\prime ,m\prime} \right)}\,\mathbb{E} \left[ e^{i\left( \phi _{n,m}-\phi _{n\prime ,m\prime} \right)} \right]}}.
\end{align}
\begin{comment}
\begin{align}
\mathbb{E}\left[\left| S \right|^2\right]=\sum_{n,m}{\sum_{n^{\prime},m^{\prime}}{e^{i\left( \alpha _{n,m}-\alpha _{n^{\prime} ,m^{\prime}} \right)}\mathbb{E}\left[e^{i\left( \phi _{n,m}-\phi _{n^{\prime} ,m^{\prime}} \right)}\right]}}. 
\end{align}
\end{comment}
Given that $\phi _{n,m}$ denotes Gaussian random variables with variance $\sigma_\phi^2$, the expectation of phase noise component can be expressed as
\begin{align}\label{25}
\mathbb{E} \left[ e^{i\left( \phi _{n,m}-\phi _{n^{\prime} ,m^{\prime}} \right)} \right] =\left\{ \begin{array}{c}
	1, n=n^{\prime} ,m=m^{\prime}\\
	\mu ^2, n\ne n^{\prime} ,m\ne m^{\prime}\\
\end{array} \right.,
\end{align}
with $\mu$ expressed as
\begin{align}\label{26}
 \mu &=\mathbb{E} \left[ e^{i\phi _{n,m}} \right] =\int_{-\infty}^{+\infty}{e^{ix}\frac{1}{\sqrt{2\pi \sigma _{\phi}}}}e^{-\frac{x^2}{2\sigma _{\phi}^{2}}}dx \nonumber
\\
&=\int_{-\infty}^{+\infty}{\frac{1}{\sqrt{2\pi \sigma _{\phi}}}}e^{-\frac{\left( x-i\sigma _{\phi}^{2} \right) -i^2\sigma _{\phi}^{4}}{2\sigma _{\phi}^{2}}}dx =e^{-\frac{\sigma _{\phi}^{2}}{2}}.
\end{align}
Substituting \eqref{26} and \eqref{25} into \eqref{24}, we obtain $\mathbb{E} [ \left| S \right|^2 ]$ as
\begin{align}
\mathbb{E} [ \left| S \right|^2 ]
&=N+\mu ^2\left( \left| \sum_{n,m}{e^{i\alpha _{n,m}}} \right|^2-N \right) \nonumber
\\
&=\mu ^2\frac{P_l}{\beta _l}+N\left( 1-\mu ^2 \right).
\end{align}
Thus, we complete the derivations of Eq. \eqref{18}.
%\clearpage
\bibliographystyle{IEEEtran}
%\balance
\bibliography{reference}

\end{document}